\documentclass[12pt,preprint]{aastex}

\newcommand{\bvf}{Brunt-V\"{a}is\"{a}l\"{a} }
\newcommand{\teffm}{$T_{\rm eff}$ }
\newcommand{\teff}{\rm T_{eff}}
\newcommand{\msunm}{$M_\odot$}

\newcommand{\pdot}{\.{P}}
\newcommand{\pdots}{\.{P}'s}

\newcommand{\massteff}{$\rm M_* - T_{eff}$ }
\newcommand{\mhmhe}{$\rm log(M_H) - log(M_{He})$ }
\newcommand{\gae}{$\rm g_{ae}$}

\begin{document}

\shorttitle{Limits on the axion mass}
\shortauthors{Bischoff-Kim et al.}

\title{Strong limits on the DFSZ axion mass with G117-B15A}
\author{A. Bischoff-Kim, M.H. Montgomery, D.E. Winget}
\affil{The University of Texas at Austin, Astronomy Department, 1 University Station, C1400, Austin,
TX 78712, USA}
\email{agnes@astro.as.utexas.edu, mikemon@astro.as.utexas.edu, 
dew@astro.as.utexas.edu}

\begin{abstract}
We compute rates of period change (\pdots) for the 215s mode in G117-B15A and 
the 213s mode in R548, first for models without axions, and then for models 
with axions of increasing mass. We use the asteroseismological models for 
G117-B15A and R548 we derived in an earlier publication . For G117-B15A, we 
consider two families of solutions, one with relatively thick hydrogen layers 
and one with thin hydrogen layers. Given the region of parameter space 
occupied by our models, we estimate error bars on the calculated \pdots \ 
using Monte Carlo simulations. Together with the observed \pdot \ for 
G117-B15A, our analysis yields strong limits on the DFSZ axion mass. Our thin 
hydrogen solutions place an upper limit of 13.5 meV on the axion, while our 
thick hydrogen solutions relaxes that limit to 26.5 meV.
\end{abstract}


\section{Astrophysical context}

G117-B15A and R548 are pulsating white dwarfs with atmospheres dominated by 
hydrogen. These stars are called DAVs or ZZ Ceti stars. Their effective 
temperatures are close to 12,000 K. They are non-radial, g-mode pulsators (the 
restoring force is buoyancy). G117-B15A is the most stable optical clock known.
The rate of change in period (\pdot = dP/dt) of its most stable mode (215s) is 
$3.57 \pm 0.82 \times 10^{-15}$ s/s \citep{keplerpdot}. That is, G117-B15A 
loses one ``tick'' every 1.7 billion years. We can use this well-measured 
\pdot \ to constrain among other things the mass of the axion, the best 
motivated dark matter candidate and in general, the emission rate of weakly
interacting particles. While measuring a \pdot \ for R548's most stable mode 
(near 213s) is less trivial because it is a doublet, the current upper limit, 
$5.5 \pm 1.9 \times 10^{-15}$ s/s \citep{Anjum0}, also indicates that in the 
near future R548 may become useful to constrain the axion mass as well. 

A \pdot ~ of order $10^{-15}$ for DAVs is consistent with a slow increase in 
period due to the cooling of the interior. It is an evolutionary timescale $-$
a measure of how fast the star is cooling. Knowing \pdot, we can determine how 
much energy a star like G117-B15A is losing every second. This energy loss 
includes of course the radiation of photons (Mestel cooling, \citealp{mestel}),
but in principle could also include the emission of weakly interacting 
particles. While neutrinos immediately come to mind, the Standard Model of
Particle Physics predicts that neutrino emission should be negligible in
G117-B15A and R548 \citep{winget04,myphdthesis}. One possibility includes 
axions and we apply our method to the latter in the present study.

Axions arise from an elegant solution to a problem with the Standard Model of 
particle physics, the strong CP problem (e.g. \citealp{axions,myphdthesis}).  
Along with supersymmetric particles, axions are currently favored candidates 
for dark matter. But they have not been discovered (neither have supersymmetric
particles) and the theory of axions fails to place any constraint on their 
mass. The possible contribution of axions to dark matter depends on their mass.
The mass of axions determines how strongly they interact with the matter we 
know, with more massive axions interacting more strongly. In turn, the stronger
the interaction of axions with matter or light, the larger their emission rate.
With pulsating white dwarfs, we can constrain the axion emission rates and 
therefore their mass. 

We determine limits on the emission of weakly interacting particles by 
comparing their cooling effect on a white dwarf model with the observed cooling
rates of that star (given by \pdot). The strength of those limits depends on 
the uncertainties involved in the measurement of the \pdot \ and in the 
modeling. This method has been used before by \citet{isern} and 
\citet{corsico01} to derive an upper mass limit for axions. While the limits 
on the axion obtained by those authors were stronger than those obtained 
through other methods (see \citealp{myphdthesis} for a review across fields of 
study), we now have the tools to greatly strengthen them. We present here the 
first study that takes into account the modeling uncertainties in a systematic 
way, allowing us to derive a strong upper mass limit for the axion. We present
our results in a way that facilitates their application to other weakly 
interacting particles. We also publish for the first time the \pdot \ we would 
expect to observe in R548 as a function of the emission rate of weakly 
interacting particles. Together with a future measurement of \pdot \ for 
R548, these results can be used to reinforce the limits given by G117-B15A.

In \citet{paper1}, thereafter paper I, we used a systematic, fine grid search 
to find best fit models to G117-B15A and R548. We were able to define the 
regions of parameter space the fits occupied. In this work, we use Monte Carlo 
simulations to calculate \pdots \ that include theoretical uncertainties, for 
varying axion mass. We begin with a brief introduction to axions in section 2. 
In section 3, we give an overview of the work done by \citet{isern} and 
\citet{corsico01}. We detail our method in section 4, discuss our results in 
section 5, and conclude in section 6.

\section{An introduction to axions}
 
The Standard Model of particle physics does not explain why Charge-Parity (CP) 
should be violated in weak interactions but not in strong interactions. One 
would therefore expect the neutron to have an electric dipole moment. 
Experiments have placed an upper limit on the neutron's electric dipole of 
$10^{-25}$ e-cm (electron charge $\times$  length of dipole), 10 orders of
magnitude below the predicted value \citep{axions}.

In more technical terms, in QCD, the neutron electric dipole moment arises from
a CP (as well as C and P) violating term in the lagrangian \citep{axions}. This
electric dipole moment is predicted to be of order 
$5 \times 10^{-16} \bar{\theta} \; \rm{ecm}$, where $\bar{\theta}$ is a free 
parameter in the theory. The experimental limit of $10^{-25} \rm ecm$ means 
that $\bar{\theta}$ is less than $10^{-10}$. Theoretical physicists do not 
understand why $\bar{\theta}$ has to be so small. This is called the strong 
CP problem and can be solved elegantly by the introduction of a new symmetry.
This symmetry, added to the Lagrangian of the fundamental interactions, is 
called the Peccei-Quinn symmetry \citep{pq}. The spontaneous breaking of this
symmetry gives rise to axions. The theory does not place any limit on their 
mass.

Axions can couple to photons, electrons (leptons) and nucleons (baryons). The 
relative strength of each coupling depends on one or more of the Peccei-Quinn 
charges of the u and d quarks and the electron (denoted $X_u$, $X_d$, and $X_e$
respectively.) Those charges can be between 0 and a number of order 1 and are 
not constrained by the theory. This gives rises to a continuum of axion models.
Two simple models include the KVSZ model \citep{kvsz1,kvsz2}, where $X_e = 0$ 
(no coupling to electrons), and the DFSZ model \citep{dfsz1,dfsz2}, where 
$X_u \sim X_d \sim X_e \sim 1$.

In this paper, we focus on DFSZ axions, those that interact with electrons. 
For a more complete review of the different axion models and their interaction
with light and matter, see \citet{myphdthesis}. In white dwarf interiors, the 
dominant axion emission mechanism would be electron bremsstrahlung, where an 
electron emits an axion as it gets accelerated in the neighborhood of an ion. 
The axion emission rate (per unit mass) for that process may be expressed as 
\citep{phonons}
\begin{equation}
\label{epsax}
\epsilon_a = 1.08 \times 10^{23} \; \rm ergs \; g^{-1} \; s^{-1}\alpha \: 
\frac{Z^2}{A} \: T^4_7 \: F(T,\rho),
\end{equation}
where  $\rm T_7 = T/10^7K$, $\alpha = g_{ae}^2/4\pi$, and $g_{ae}$, the
strength of the axion-electron coupling is given by equation \ref{gae} below. 
$\rm F(T,\rho)$ is a numerical fit provided by \citet{phonons}. It is of order 
1 throughout most of the interior of a typical white dwarf model (e.g. figure
\ref{f1}).

The strength of the axion-electron coupling is 
\begin{equation}
\label{gae}
\rm g_{ae} = (2.8 \times 10^{-11}) \; \frac{m_a \cos^2\beta}{1 \; 
\rm{eV}}, 
\end{equation}
where $\rm m_a$ and $\beta$ are free parameters. For DFSZ axions, the coupling 
to electrons is $10^8$ orders of magnitude greater than the coupling to photons
(assuming $\rm cos^2\beta \sim 1$). Following in \citet{isern}'s footsteps, we 
shall not make make any assumptions on the value of $\beta$ and simply state 
that the quantity we are constraining is $\rm m_a \cos^2\beta$. We shall, 
however, refer to it simply as the ``axion mass''.

\section{Early work}

In a pioneering work, \citet{isern} used G117-B15A's \pdot \  to obtain a limit
on the axion mass. At the time, the \pdot \  measured ($12.0 \pm 3.5 \times 
10^{-15} \rm s/s$, \citealp{keplerpdot1}) for that star was uncertain, and much
higher than the one expected from simple Mestel cooling. Using models available
at the time \citep{wood90,dantona} and a simple semi-analytical treatment, 
\citeauthor{isern} found an average axion mass of 8 meV. Individual values, 
depending upon the model chosen and value of observed \pdot \  considered 
(\pdot \ $-\; \Delta$\pdot, \pdot, and \pdot \  + $\Delta$\pdot) allowed a 
range between 0 meV and 20 meV for the axion mass.

\citet{corsico01} revisited the problem with a new measured value of \pdot, 
that had since decreased to what was expected from simple Mestel cooling 
\citep{mestel}: $(2.3 \pm 1.4) \times 10^{-15} \rm s/s$ \citep{keplerpdot2}. In
their work, \citeauthor{corsico01} performed an asteroseismological study of 
G117-B15A to find its mass, helium layer mass, and hydrogen layer mass. To 
reduce the number of parameters to fit, they fixed the internal composition to 
that found from stellar evolution calculations by \citet{salaris}, and the 
effective temperature to the latest spectroscopic estimate at the time, 
11,620 K \citep{b95a}. Their best fit model had a mass of 0.55 \msunm, a helium
layer mass $\rm M_{He}=10^{-2} M_*$ and a hydrogen layer mass $\rm M_{He}=
10^{-4} M_*$. On the average, the periods of that model are 5 seconds away 
from the observed periods. In contrast, the models we use in the present study 
fit to better than 1 second.

\citet{corsico01} considered only small uncertainties in effective temperature 
(200K) and found that they led to a 4\% uncertainty in the calculated \pdots. 
They also considered larger uncertainties in mass and central oxygen abundance 
and found that those had less of an effect on the model's \pdots \  than the 
effective temperature. A mass uncertainty of .02 \msunm \ (3\%) led to a 6\% 
uncertainty in \pdots. And considering a full range of core composition (0\% 
carbon to pure carbon) changed the \pdots \  only by 5\%. They concluded that 
uncertainties other than the one in the measured \pdot \  were insignificant 
and could be ignored altogether. They found a tight constraint on the 
axion mass: 4 meV. If we adopt Corsico et al.'s best fit parameters for 
G117-B15A, and follow the same method using our models, we obtain a similar 
limit on the axion mass.

To place things in prospective, the best DFSZ axion mass limits aside from that
of \citet{corsico01} and the ones presented here come from the observation of 
Red Giants and Horizontal Branch stellar populations in clusters. If axions 
were massive enough, they would cool the core of stars and have an observable 
effect on the morphology of those populations. The best limit found using this 
method is 9 meV \citep{RG4}. Such limits depend on the assumed core mass in the
models, a rather uncertain quantity. The pulsating white dwarf method does not
suffer from such large uncertainties.

Since the work done by \citet{corsico01}, a newer value of \pdot \ = 
$3.57 \pm 0.82 \times 10^{-15} \rm s/s$ has been obtained \citep{keplerpdot}. 
The new value has a smaller measurement error, and could help constrain the 
axion mass better. More importantly, we take into account additional modeling
uncertainties that have been ignored in previous analyses. The better we 
account for those uncertainties, the stronger our limit on the axion mass will 
be. We will see that this leads us to different conclusions than the authors 
mentioned above.

\section{Method}

\subsection{Axion emission in our models}

For a mixture of carbon and oxygen, as in the interior of our white dwarf
models, the axion emission rate is given by
\begin{equation}
\label{epsmixed}
\epsilon_a \rm = X_c \epsilon_c + X_o \epsilon_o , 
\end{equation}
where $\rm X_c$ and $\rm X_o$ are the carbon and oxygen abundance, and 
$\rm \epsilon_c$ and $epsilon_o$ are the axion emission rates in pure
carbon and pure oxygen respectively.  Combining equation \ref{epsax} and 
\ref{epsmixed} we have:
\begin{equation}
\label{epsmixed2}
\epsilon_a = 1.08 \times 10^{23} \; \alpha \: T^4_7  \left[3X_cF_c(\rho,T) + 
4X_oF_o(\rho,T) \right].
\end{equation}
We distinguish $\rm F(T,\rho)$ in pure carbon and pure oxygen by using 
subscripts.

\begin{figure}[!ht]
\begin{center}
\rotatebox{270}{\scalebox{0.5}{\includegraphics{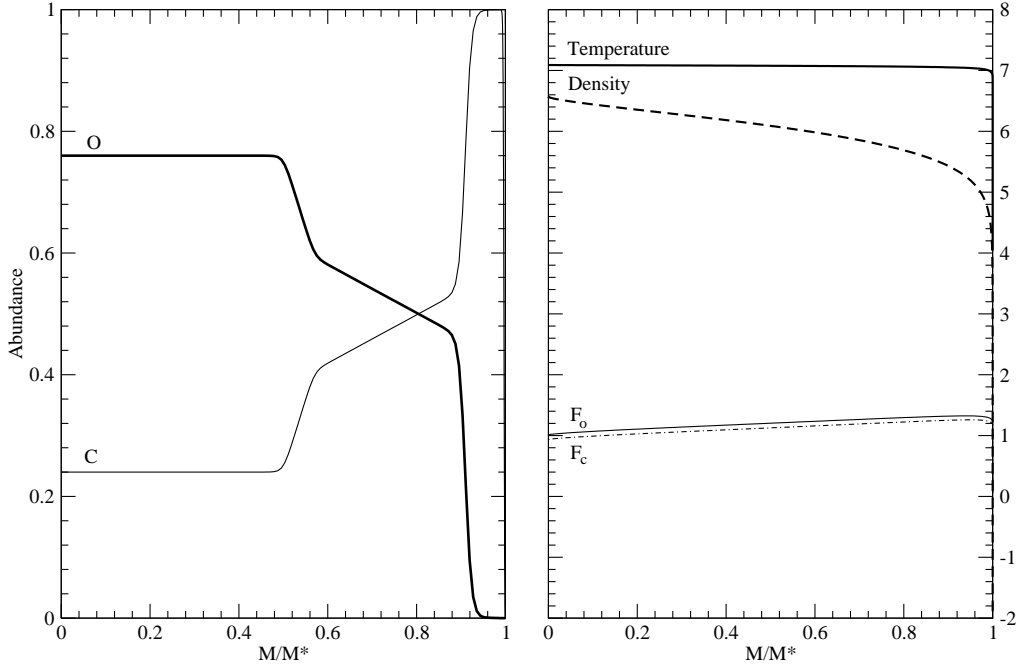}}}
\end{center}
\caption{
\label{f1}
Internal properties of the fiducial model. Left panel: carbon and oxygen 
abundance. Right panel: log of the temperature in K, log of the density in
g/cc, $\rm F_o$ and $\rm F_c$ (see text).}
\end{figure}

Integrating over the mass of the model, we obtain the axion luminosity. 
While we carry the exact integration numerically in our models, it is
useful at this point to write down an approximate expression for the 
axion luminosity. We use the internal properties of a fiducial model, chosen
because it matches G117-B15A's periods and \pdot \ for the 215s mode very 
closely. This fiducial model has \teffm = 12656 K, $M_*$ = 0.602 \msunm, a
helium layer mass of $3.55 \times 10^{-3} M_*$, and a hydrogen layer mass
of $4.79 \times 10^{-8} M_*$. We display the internal properties of the 
fiducial model in figure \ref{f1}.

\clearpage

Following the example set forth in \citet{isern}, we make the following 
approximations: 1) We set $\rm T = T_{center} = 1.2 \times 10^7 K$ throughout 
the model; 2) We also set $\rm F_c$ and $\rm F_o$ constant 
($\rm F_c \approx 1.1$ and $\rm F_o \approx 1.2$). The axion luminosity is
then
\begin{equation}
\rm L_a = 1.08 \times 10^{23} \; \alpha \: T^4_7 \: M_* \left(3F_c\int_{0}^{1}
\rm X_c \, dm +  4F_o\int_{0}^{1}\rm X_o \, dm \right),
\end{equation}
where $\rm m\equiv M_r/M_*$. For the fiducial model's core composition, 
$\int_{0}^{1} \rm X_c \, dm \approx 0.4$ and $\int_{0}^{1}\rm X_o \, dm
\approx 0.6$. We obtain
\begin{equation}
\label{apprxla}
\rm L_a = 1.13 \times 10^{57} \; \alpha \; \rm ergs \; s^{-1}.
\end{equation}
This approximate expression is 7\% accurate for our fiducial model.

\subsection{Asteroseismological fits}

Before we include axions in our models, we need to make sure we have reliable 
models of the stars under study, G117-B15A and R548. We performed our own 
asteroseismological analysis of those stars in paper I. We varied 4 parameters,
including the effective temperature and the helium layer mass, $\rm M_{He}$ (in
addition to the stellar mass and the hydrogen layer mass, $\rm M_H$). In paper 
I, we explored systematically a number of sources of uncertainties in our 
models and concluded that we could adequately account for them by treating 
equally all models that matched the observed periods to better than 1 second. 
With that cut-off, we defined the regions of parameter space our models 
occupied. We found that our best fit models were hot and/or massive compared to
the spectroscopy, though they were not completely inconsistent with the latter.
Work is in progress to resolve this apparent discrepancy. Our calculations are
self-consistent and we proceed to derive meaningful results.

\clearpage

\begin{figure}[!h]
\begin{center}
\scalebox{0.4}{\includegraphics{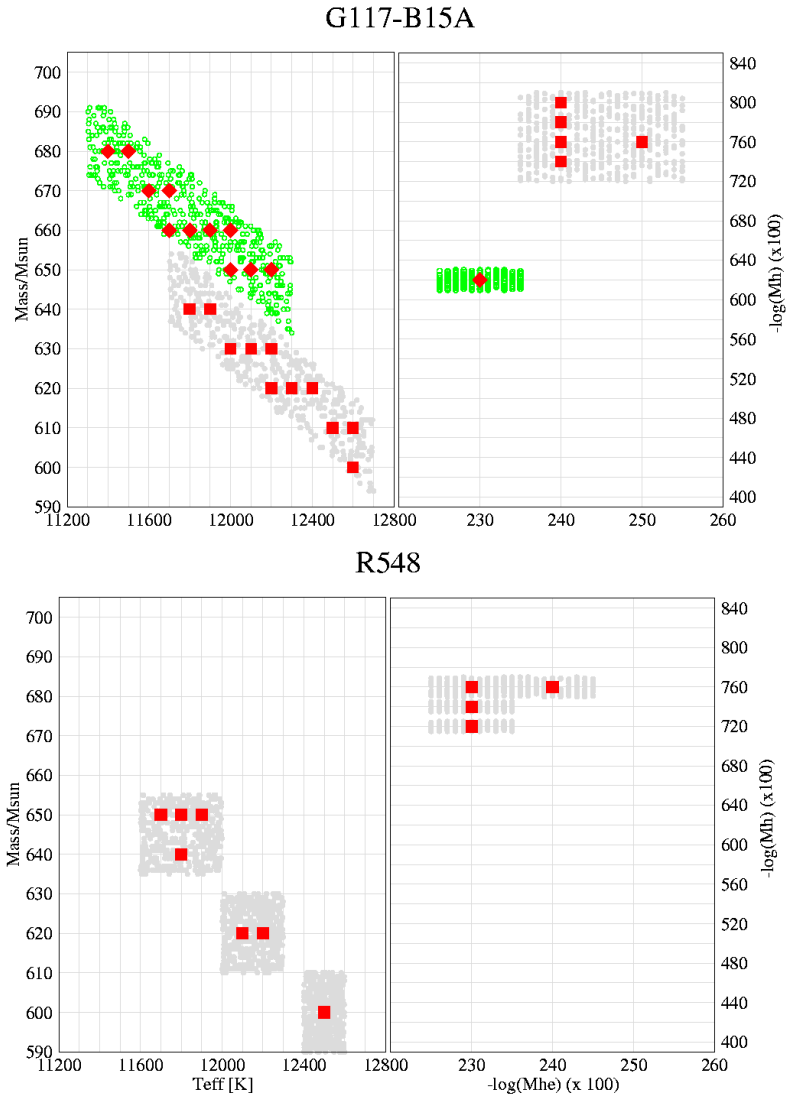}}
\end{center}
\caption{
\label{samplepopg117}
A sample simulated random population for G117-B15A and R548 in the \massteff 
plane and \mhmhe plane. The small symbols show the location in parameter space 
of the simulated population, while the squares and diamonds are best fit grid 
models from paper I.}
\end{figure}

\clearpage

\subsection{Monte Carlo Simulations}

Having narrowed down the regions of parameter space occupied by best fit 
models for G117-B15A and R548, the next step is to derive rates of period 
change for those models along with the associated uncertainties. This is a 
classic Monte Carlo problem. For G117-B15A, we chose $\teff$, $\rm \Delta M$ 
(instead of $\rm M_*$), $\rm M_{He}$, and $\rm M_H$ as our 4 independent 
variables, where $\rm \Delta M$ is defined by
\begin{equation}
\label{mteff}
M_* = b - a \; \teff + \Delta M, 
\end{equation}
and $a$ and $b$ are defined so that the simulated populations occupy the 
region of parameter space in the \massteff plane defined by the results from
our astereseismological fits. For each family of solutions we found in paper I,
without including axions just yet, we generated a random population (N=500). We
display the distribution of the models of each of the populations in figure
\ref{samplepopg117}.

We have two possible families of fits for G117-B15A. One has a thicker hydrogen
layer ($\rm M_H = 6.3 \times 10^{-7} M_*$) and occupies the high mass, low 
effective temperature region of the \massteff plane. The other has a thin 
hydrogen layer ($\rm M_H < 6.3 \times 10^{-7} M_*$) and occupies the higher 
effective temperature (lower mass) region of the \massteff plane. The thin 
hydrogen fits to G117-B15A are very similar to the fits to R548. While this 
makes the thin hydrogen fits more attractive than the thick hydrogen fits (by 
``thick'' we mean ``not as thin''), we cannot discard the latter as a viable 
possibility on that basis alone. We shall consider both families when 
constraining the axion mass.

\begin{table}
\caption{
\label{means}
Average periods and \pdots \  for G117-B15A and R548, without axions.}
\smallskip
\begin{center}
\begin{tabular}{|l|cc|c|}
%\tableline
\hline
\rule[-0.3cm]{0mm}{0.6cm}
              & \multicolumn{2}{|c|}{G117-B15A} & R548                       \\
\hline
\rule[-0.0cm]{0mm}{0.6cm}
$\rm M_H$   & $6.3 \times 10^{-7}$ & $4 \times 10^{-8}$ & $4 \times 10^{-8}$ \\
\hline
\rule[-0.0cm]{0mm}{0.6cm}
P (observed)   [s] & \multicolumn{2}{|c|}{215.197389} & 213                  \\
~P (calculated) [s]& $214.9 \pm 2.5$ & $215.7 \pm 2.0$ & $213.1 \pm 1.8$     \\
\hline
~\pdot \ (observed) $[10^{-15} \rm s/s]$
                   & \multicolumn{2}{|c|}{$3.57 \pm 0.82$} & $<5.5 \pm 1.9$  \\
\rule[-0.3cm]{0mm}{0.6cm}
\pdot \ (calculated)  $[10^{-15} \rm s/s]$ & 
                $1.92 \pm 0.26$    & $2.98 \pm 0.17$    & $2.91 \pm 0.29$    \\
\hline
\end{tabular}
\end{center}
\end{table}

For each model in each population, we proceeded with calculating a \pdot. We 
checked that the period and \pdot \  distribution for each of the populations 
were close to Gaussian and found that they were. For each population (family 
of fits) we calculated the mean period and \pdot. We list the results in 
table \ref{means} with one sigma error bars. Remember that these results assume
axions do not exist and that the cooling is due entirely to photons. For 
G117-B15A, the thick hydrogen fit \pdot \ is two sigma below the observed 
value. However, we cannot discard this solution based on a low \pdot \ alone, 
as we can always increase the cooling rate and raise the value of \pdot \ by 
adding axion emission to the models.

Next we included axion production in our models and repeated the procedure 
for each different axion mass. We confirm an important fact \citet{corsico01}
first pointed out: the periods themselves are insensitive to the axion mass
(for $\rm m_a$ up to 30 meV at least). The reason for this small dependence is 
that axion emission does not introduce any "bumps" into the \bvf frequency. It 
merely slightly lowers it throughout the degenerate core. This allows us to
meaningfully compare the calculated \pdots \ with the observed value.

\section{Results}

\begin{figure}[!ht]
\begin{center}
\rotatebox{270}{\scalebox{0.5}{\includegraphics{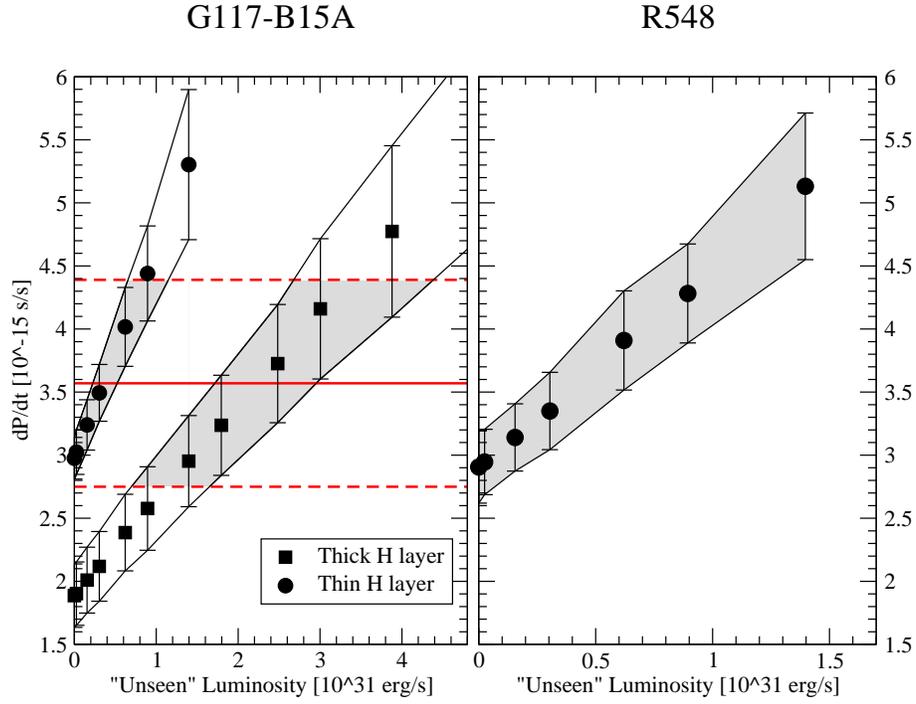}}}
\end{center}
\caption{
\label{pdotmax}
\pdots \  as a function of the luminosity due to the emission of 
weakly interacting particles (in that case, axions) for G117-B15A and R548. 
The observed \pdot \  for G117-B15A is indicated by the horizontal solid line 
and its 1 sigma error bars by the dashed lines above and below it. The shaded 
regions inside the dashed horizontal lines indicates the range of calculated 
\pdots \ consistent with the observed value (to 1 sigma).}
\end{figure}

In figure \ref{pdotmax} we show \pdot \  as a function of the axion luminosity 
for G117-B15A and R548. While we included axions specifically in our models,
the \pdots \ depend only on the total luminosity, not on the exact nature of
the weakly interacting particles responsible for the ``unseen'' loss of energy.
Our results may therefore directly be generalized to weakly interacting 
particles other than axions.

The thin hydrogen solution for G117-B15A places an upper limit of $1.1 \times
10^{-31}$ erg/s for the energy loss rate due to weakly interacting particles.
For axions, this translates to an upper limit of $3.78 \times 10^{-13}$ on 
the axion-electron coupling constant \gae \ (13.5 meV on the axion mass).The 
thick hydrogen family of fits (taken alone) constrains the ``unseen''
luminosity to be between $2.7 \times 10^{-31}$ and $4.4 \times 10^{-31}$ 
erg/s. These limits translate to $2.9 \times 10^{-13} \leq$ \gae \ $\leq 7.4
\times 10^{-13}$ (axions between 10.4 and 26.5 meV in mass). Combining the two 
families together, we can place a conservative upper limit of 
$4.4 \times 10^{-31}$ erg/s on the luminosity of weakly interacting particles
that would be emitted in the interior of G117-B15A, and a 26.5 meV upper limit
on the axion mass. For R548, we are still awaiting a better determined \pdot \ 
to allow us to obtain a potentially tighter constraint on the emission of 
weakly interacting particles in white dwarfs.

\clearpage

\clearpage

\section{Conclusions}

We computed \pdots \  for the 215s mode in G117-B15A and the 213s mode in R548 
first for models without axions, and then for models with axions of increasing 
mass. The thin hydrogen, zero axion models for G117-B15A have a \pdot \ of 
$2.98 \pm 0.17 \times 10^{-15} \rm s/s$, consistent with the observed value of 
$3.57 \pm 0.82 \times 10^{-15} \rm s/s$. The zero axion, thick hydrogen models 
have a \pdot \ of $1.92 \pm 0.26 \times 10^{-15} \rm s/s$, 2 sigma below
the observed \pdot \ and require additional cooling through axions to become
consistent with that observed value. The zero axion models for R548 have a 
\pdot \ similar to the thin hydrogen \pdot \ for G117-B15A ($\sim 3 \times 
10^{-15} \rm s/s$), consistent with the best current limit on \pdot \  for 
R548's 213s mode. 

If we believe that G117-B15A must be similar to R548, then we find an upper 
limit on the axion luminosity of $1.1 \times 10^{-31}$ erg/s, corresponding to 
an axion-electron coupling constant \gae \ $=3.78 \times 10^{-13}$ , or an
axion mass of 13.5 meV. A second family of fits with thicker hydrogen layers
relaxes that limit to $L_{a} \leq 4.4 \times 10^{-31}$ erg/s, \gae \ $\leq 7.4
\times 10^{-13}$ or $\rm m_a \leq $ 26.5 meV. What makes this limit 
unique is not the fact that it is lower than previous limits. What makes it 
unique is its strength. It is based on an unbiased exploration of the 
parameter space of the models. Compared to typical axion searches (see 
\citealp{myphdthesis} for a review), the present analysis involved well-known
physics and relatively small, well-identified uncertainties, considered 
systematically. Improved asteroseismological models will help us choose 
between the two families of best fits for G117-B15A and improve this limit.

An even longer data baseline for G117-B15A will reduce the size of the error 
bars on the measured \pdot \ and further narrow the allowed range for the axion
mass. We can soon expect a useful \pdot \ for R548 as well, which will 
reinforce the results of this study. As a part of his planet search around 
pulsating white dwarfs,  Mullally has determined upper limits of $10^{-13}$ s/s
for an additional 9 white dwarfs \citep{fergalphdthesis} and has identified 
more blue edge DAVs with stable modes. With Argos \citep{argos}, the high-speed
photometry data collected at McDonald observatory as a part of the on-going
observation of hot DAVs should begin to yield measurements of \pdots \ for 
those stars within the next 3 to 5 years. It is merely a question of time until
we can place an even tighter constraint on the emission of weakly interacting
particles in these stars.

In the present study, we assumed that any unseen source of energy loss was due
to axions. The results we present can readily be used to constrain the 
emission of other weakly interacting particles possibly produced inside white 
dwarfs, such as supersymmetric particles.

\section{Acknowledgment}

The authors would like to thank E.L. Robinson for key suggestions that helped 
us improve this work and the referee for ideas to augment its impact. This 
research was supported by NSF grant AST-0507639. 

\bibliographystyle{apj}

\end{document}